\documentclass[%
preprint,
reprint,
superscriptaddress,
showpacs,
showkeys,
nofootinbib,
 amsmath,amssymb,
 aps,
]{revtex4-1}

\usepackage{graphicx}
\usepackage[T1]{fontenc} 
\usepackage{epstopdf} 
\usepackage{dcolumn}
\usepackage{bm}
\usepackage{hyperref}
\usepackage[mathlines]{lineno}
\usepackage{xfrac} 
\usepackage{color}

\usepackage[ulem=normalem]{changes}

\begin{document}
\preprint{APS/123-QED}

\title{Evidence of a near-threshold resonance in $^{11}$B relevant to the $\beta$-delayed proton emission of $^{11}$Be.} 

\author {Y.~Ayyad}
\email{e-mail: yassid.ayyad@usc.es }
\affiliation{IGFAE, Universidade de Santiago de Compostela, E-15782, Santiago de Compostela, Spain}
\affiliation{Facility for Rare Isotope Beams, Michigan State University, East Lansing, MI 48824, USA}
\author {W.~Mittig}
\affiliation{Facility for Rare Isotope Beams, Michigan State University, East Lansing, MI 48824, USA}
\affiliation{Department of Physics and Astronomy, Michigan State University, East Lansing, Michigan 48824, USA}
\author {T.~Tang}
\affiliation{Facility for Rare Isotope Beams, Michigan State University, East Lansing, MI 48824, USA}
\author {B.~Olaizola}
\affiliation{ISOLDE-EP, CERN, CH-1211 Geneva 23, Switzerland}
\author {G.~Potel}
\affiliation{Lawrence Livermore National Lab., P.O. Box 808, Livermore, CA 94550}
\author {N.~Rijal}
\affiliation{Facility for Rare Isotope Beams, Michigan State University, East Lansing, MI 48824, USA}
\author {N.~Watwood}
\affiliation{Facility for Rare Isotope Beams, Michigan State University, East Lansing, MI 48824, USA}
\author {H.~Alvarez-Pol}
\affiliation{IGFAE, Universidade de Santiago de Compostela, E-15782, Santiago de Compostela, Spain}
\author {D.~Bazin}
\affiliation{Facility for Rare Isotope Beams, Michigan State University, East Lansing, MI 48824, USA}
\affiliation{Department of Physics and Astronomy, Michigan State University, East Lansing, Michigan 48824, USA}
\author {M.~Caamaño}
\affiliation{IGFAE, Universidade de Santiago de Compostela, E-15782, Santiago de Compostela, Spain}
\author {J.~Chen}
\affiliation{Physics Division, Argonne National Laboratory, Argonne, IL 60439, USA}
\author {M.~Cortesi}
\affiliation{Facility for Rare Isotope Beams, Michigan State University, East Lansing, MI 48824, USA}
\author {B.~Fernández-Domínguez}
\affiliation{IGFAE, Universidade de Santiago de Compostela, E-15782, Santiago de Compostela, Spain}
\author {S.~Giraud}
\affiliation{Facility for Rare Isotope Beams, Michigan State University, East Lansing, MI 48824, USA}
\author {P.~Gueye}
\affiliation{Facility for Rare Isotope Beams, Michigan State University, East Lansing, MI 48824, USA}
\affiliation{Department of Physics and Astronomy, Michigan State University, East Lansing, Michigan 48824, USA}
\author{S.~Heinitz}
\affiliation{Laboratory of Radiochemistry, Paul Scherrer Institute, Villigen, Switzerland}
\author {R.~Jain}
\affiliation{Facility for Rare Isotope Beams, Michigan State University, East Lansing, MI 48824, USA}
\affiliation{Department of Physics and Astronomy, Michigan State University, East Lansing, Michigan 48824, USA}
\author {B.~P.~Kay}
\affiliation{Physics Division, Argonne National Laboratory, Argonne, IL 60439, USA}
\author {E.~A.~Maugeri}
\affiliation{Laboratory of Radiochemistry, Paul Scherrer Institute, Villigen, Switzerland}
\author {B.~Monteagudo}
\affiliation{Facility for Rare Isotope Beams, Michigan State University, East Lansing, MI 48824, USA}
\author {F.~Ndayisabye}
\affiliation{Facility for Rare Isotope Beams, Michigan State University, East Lansing, MI 48824, USA}
\affiliation{Department of Physics and Astronomy, Michigan State University, East Lansing, Michigan 48824, USA}
\author {S.~N.~Paneru}
\affiliation{Facility for Rare Isotope Beams, Michigan State University, East Lansing, MI 48824, USA}
\author {J.~Pereira}
\affiliation{Facility for Rare Isotope Beams, Michigan State University, East Lansing, MI 48824, USA}
\author {E.~Rubino}
\affiliation{Facility for Rare Isotope Beams, Michigan State University, East Lansing, MI 48824, USA}
\author {C.~Santamaria}
\affiliation{Facility for Rare Isotope Beams, Michigan State University, East Lansing, MI 48824, USA}
\author {D.~Schumann}
\affiliation{Laboratory of Radiochemistry, Paul Scherrer Institute, Villigen, Switzerland}
\author {J.~Surbrook}
\affiliation{Facility for Rare Isotope Beams, Michigan State University, East Lansing, MI 48824, USA}
\affiliation{Department of Physics and Astronomy, Michigan State University, East Lansing, Michigan 48824, USA}
\author {L.~Wagner}
\affiliation{Facility for Rare Isotope Beams, Michigan State University, East Lansing, MI 48824, USA}
\author {J.~C.~Zamora}
\affiliation{Facility for Rare Isotope Beams, Michigan State University, East Lansing, MI 48824, USA}
\author{V.~Zelevinsky}
\affiliation{Facility for Rare Isotope Beams, Michigan State University, East Lansing, MI 48824, USA}
\affiliation{Department of Physics and Astronomy, Michigan State University, East Lansing, Michigan 48824, USA}

\date{\today}

\begin{abstract}

A narrow near-threshold proton-emitting resonance (E$_{x}$ = 11.4~MeV, $J^{\pi}=1/2^{+}$ and $\Gamma_p$ = 4.4~keV) was directly observed in $^{11}$B via proton resonance scattering. This resonance was previously inferred in the $\beta$-delayed proton emission of the neutron halo nucleus $^{11}$Be. The good agreement between both experimental results serves as a ground to confirm the existence of such exotic decay and the particular behavior of weakly bound nuclei coupled to the continuum. $R$-matrix analysis shows a sizable partial decay width for both, proton and $\alpha$ emission channels. 

\end{abstract}

\keywords{resonance scattering, near-threshold resonance, open quantum system, $^{10}$Be, $^{11}$Be, $^{11}$B}

\maketitle

Loosely bound atomic nuclei can be understood as open quantum systems: a weakly bound ensemble of nucleons coupled to an external environment. The behavior and properties of such systems are deeply affected by the interplay with this external environment called continuum. Because of this interplay, these systems display generic properties that are common to all weakly bound systems, with near or above threshold excitation energies. This coupling to the continuum may manifest in a nuclear reaction that excites the system to a state near particle emission threshold. Therefore, the study of these atomic quantum systems underlines the, commonly contrived, close link between reaction and structure. As the system becomes gradually less bound, many-nucleon correlations may manifest through the formation of particle clusters via narrow resonances in the vicinity of the particle emission threshold. Although the formation and emission of clusters with well defined quantum states is ubiquitous in the nuclear physics domain, little is known about how their properties are defined.\\ 

Many examples of particle-emitting near-threshold narrow resonances of fundamental relevance for $\alpha$ clustering~\cite{10.1143/PTPS.E68.464}, proton radioactivity~\cite{girardalcindor2021narrow}, and for reactions of astrophysical interest~\cite{doi:10.1142/10429,PhysRevC.101.052802} can be found throughout the entire nuclear landscape. Such a correlation-driven nuclear binding gives rise to open quantum systems where the radial wave function of valence nucleons extends well beyond the bound core forming weakly-bound nuclei known as halo~\cite{Tanihata_1996}. Open quantum systems with particle emitting states can be formed near the drip line where separation energies become negative~\cite{PhysRevC.105.014314}, $\beta$-decay into unbound states~\cite{PhysRevC.93.065803}, or by resonance scattering~\cite{DEGRANCEY2016}. A very particular, near-threshold, narrow resonance was recently inferred from the $\beta$-delayed proton emission ($\beta$p) of the halo nucleus $^{11}$Be, a counter-intuitive decay in neutron-rich nuclei~\cite{Ayyad2019,Ayyad2019Err}. This exotic decay is possible, within a relatively small energy window, for systems with a low neutron separation energy, such as $^{11}$Be (501.6(3)~keV)~\cite{JONSON200177}. One of the key questions is whether this type of decay proceeds via a two-step mechanism feeding unbound states on the daughter nucleus, or directly into the continuum. There exists clear experimental evidence supporting a direct decay in the $\beta$-delayed deuteron emission of $^{6}$He~\cite{PhysRevC.80.054307} and of $^{11}$Li~\cite{PhysRevLett.101.212501}. Branching ratios for these decays amount to the order of 10$^{-6}$ and 10$^{-4}$, respectively.

The $\beta$p decay of $^{11}$Be was directly observed for the first time in the $^{11}\text{Be}\rightarrow ^{10}\text{Be} + \beta^- + p$ disintegration by our collaboration~\cite{Ayyad2019}. The experiment was performed by implanting a $^{11}$Be beam in the prototype Active Target Time Projection Chamber (pATTPC)~\cite{Suzuki2012}. A novel particle tracking algorithm was employed to distinguish between protons, $\alpha$ particles and recoiling nuclei. The experiment yielded a branching ratio for the $\beta^-\text{p}$ branch of b$_p = 1.3(3) \cdot 10^{-5}$. Moreover, from the energy distribution of the emitted protons, it was deduced that the $\beta^-\text{p}$ decay was sequential. The disintegration proceeded via an intermediate near-threshold narrow resonance in the $\beta^-$ decay product $^{11}\text{B}^*$ at an energy E$_R = 196(20)$~keV above the proton separation energy and a total width of $\Gamma_p = 12(5)$~keV and $J^{\pi}=(1/2^{+},3/2^{+}$). No corresponding state had been observed in $^{11}$B at the time. This result was contested by Riisager and collaborators who employed an indirect approach to measure this same branching ratio~\cite{Borge2013a, Riisager2014a, Riisager2020}. They made use of a mass separator to measure the presence of $^{10}$Be in a catcher where $^{11}$Be had been implanted. Despite improving their experiment on at least three occasions, unexplained discrepancies between their results persisted. Due to this inconsistencies, an upper limit was adopted from the lowest branching ratio, b$_p < 2.2 \cdot 10^{-6}$~\cite{Riisager2020}, in clear disagreement with the value reported in Ref.~\cite{Ayyad2019} and with their previous measurement~\cite{Riisager2014a}.\\

These two experiments also linked $^{11}$Be to the search for the decay of neutrons into dark matter, the so called dark decay~\cite{PhysRevLett.120.191801}. The dark decay, that involves physics beyond the Standard Model, tries to explain the long-standing neutron lifetime puzzle, by hypothesizing that $\sim 1\%$ of free neutrons decay into an undetected dark-sector particle instead of a proton. In this model, weakly-bound neutrons could also undergo dark decay, with the halo neutron in $^{11}$Be being the most promising candidate~\cite{PhysRevC.97.042501}. The final product of the $^{11}$Be dark decay would be a $^{10}$Be nucleus plus an undetected dark particle. A precise measurement of the $^{11}$Be$\rightarrow ^{10}$Be rate (similar to the attempts by Riisager and collaborators~\cite{Borge2013a, Riisager2014a, Riisager2020}) would measure a combination of the $\beta p$ and dark decay branching ratios. It is, therefore, paramount to have a precise measurement of the $\beta p$ mechanism and branching ratio in order to disentangle its contribution to the overall $^{11}$Be$\rightarrow ^{10}$Be decay and thus extract any hypothetical dark decay branch.\\

There have been several attempts from theory to confirm the resonance in $^{11}$B and to estimate the $\beta^-\text{p}$ decay branching ratio in $^{11}$Be. Before the experiments were conducted, Baye and Tursunov~\cite{Baye2011} deduced b$_p\sim 5 \cdot 10^{-9}$ employing a cluster model with no resonant intermediate state. More theoretical attempts were carried out after the publication of the first experimental $\beta^-\text{p}$ results. Volya~\cite{Volya2020} performed shell model calculations and concluded that no suitable resonance existed in $^{11}$B that could act as an intermediate state in order to enhance the $\beta^-\text{p}$ decay branching ratio. These calculations also suggested that, if such a state existed, it would strongly favor breaking into $^{7}$Li$+\alpha$ rather than emitting a proton. Oko\l{}owicz and collaborators~\cite{Okolowicz2020} arrived at a rather different conclusion; using a shell model embedded in the continuum model (SMEC), they were able to infer the presence of the intermediate resonance with $J^{\pi}=1/2^{+}$ and a small contribution from the $^{7}$Li+$\alpha$ channel which highlights the orthogonality of both possible eigenstates. However, their model does not reconcile with the large b$_p$ obtained in Ref.~\cite{Ayyad2019}. Their most recent calculations suggest that the b$_p$ should be 40 times lower to harmonize with the $\Gamma_p$ and the branching ratio for $\alpha$ decay (b$_\alpha$)~\cite{okolowicz2021betarm}. In their study, it is assumed that there exists a very close 3/2$^{+}$ resonance (11.49~MeV) that decays predominantly by $\alpha$ emission. Such a resonance was indirectly deduced from an R-matrix fit but never observed before~\cite{PhysRevC.99.044316}. They also conclude that decay from the Isobaric Analogue State (IAS), as suggested by Ref.~\cite{Volya2020}, is ruled out. Lastly, Elkamhawy \textit{et al.}~\cite{Elkamhawy2021} performed Halo Effective Field Theory with and without the intermediate resonance state in $^{11}$B. Similarly to Ref.~\cite{Baye2011}, for a direct decay (no resonance) the b$_p$ obtained was orders of magnitude lower than the directly measured one~\cite{Ayyad2019}. On the other hand, when a resonance with parameters similar to those measured by this collaboration was introduced, the experimental b$_p$ was reproduced.\\

\begin{figure}
\begin{center}
\includegraphics[width=\columnwidth, keepaspectratio]{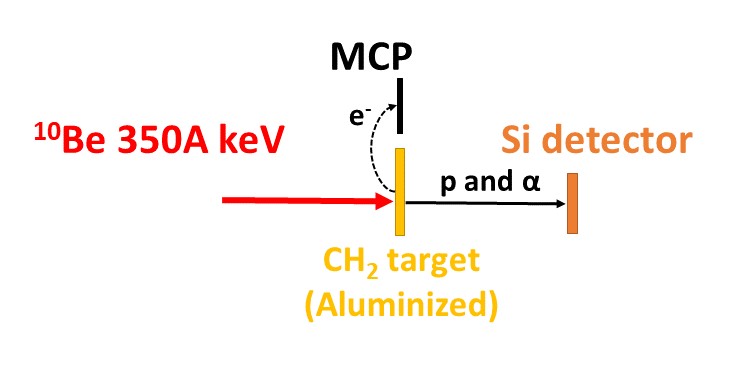}
\includegraphics[width=\columnwidth, keepaspectratio]{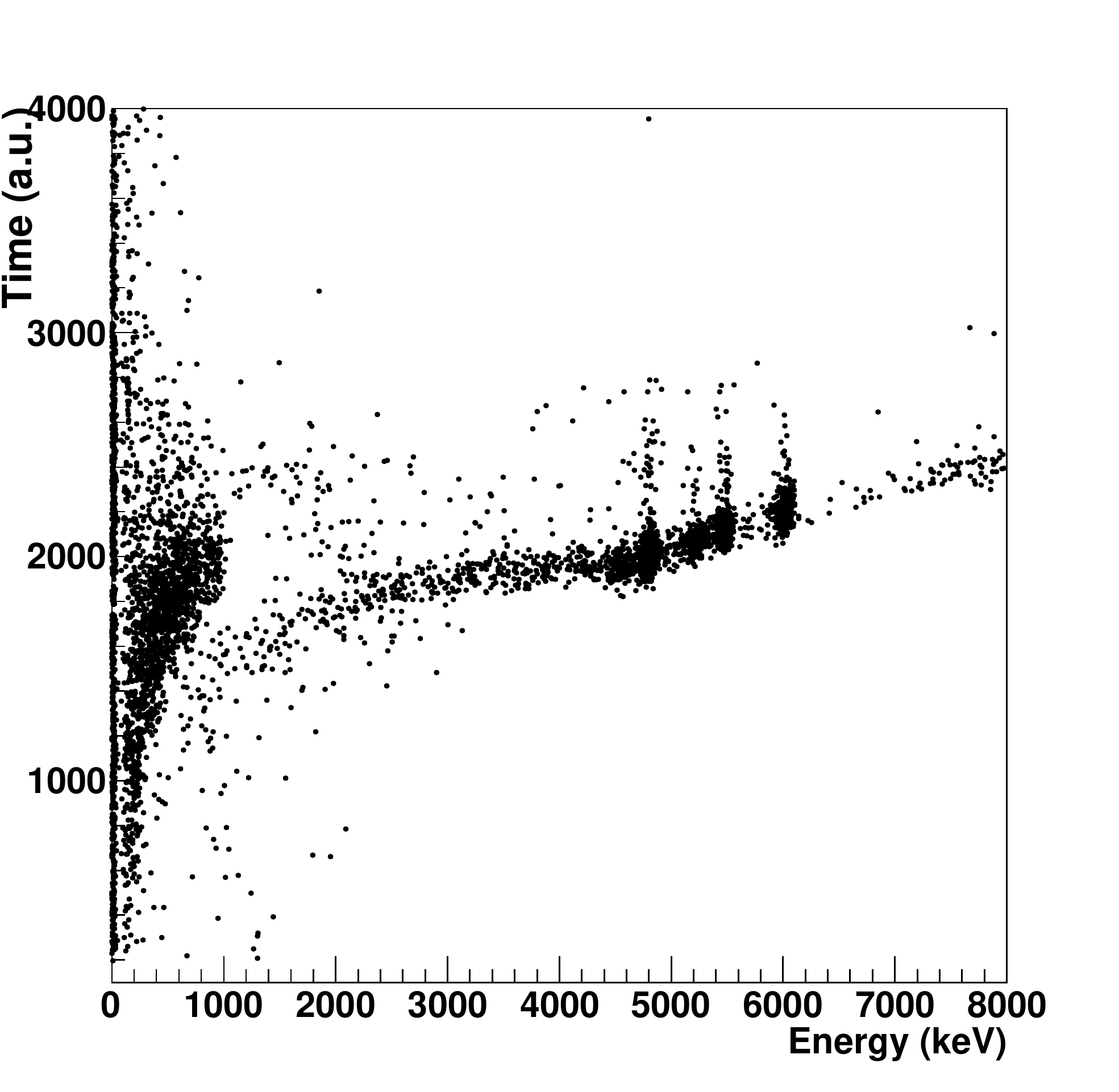}
\caption{\label{fig:fig_1} Upper panel: Sketch of the experimental setup. Lower panel: Particle (proton and $\alpha$) energy vs Time-of-Flight. The peaks at high energy correspond to the $^{228}$Th source $\alpha$ particles. }
\end{center}
\end{figure}

It is clear, thus, that the exotic $\beta^-\text{p}$ decay requires the presence of a near-threshold resonance to enhance it to the level observed in Ref.~\cite{Ayyad2019}. Since no suitable level has been observed in $^{11}$B to date, a dedicated experiment employing the $^{10}$Be(p,p) reaction was performed to clarify its existence and properties. The experiment was conducted at the ReA3 re-accelerator facility of the National Superconducting Cyclotron Laboratory (NSCL) using a pure 350$A$~keV $^{10}$Be beam with an intensity of about 10$^{3}$ pps. The $^{10}$Be material was produced at Paul Scherrerr Institut (Villigen, Switzerland) from proton-irradiated carbon~\cite{HEINITZ2017260}. The excitation function of the reaction was obtained by stopping the beam on a 9.6~$\mu$m thick CH$_{2}$ target foil (8.64~mg/cm$^{2}$). A very thin (tens of nm) aluminum layer was evaporated on the upstream side of the foil. Secondary electrons produced from the aluminum by the beam were deflected using permanent magnets into a Microchannel Plate Detector in Chevron mode manufactured by TECTRA. A 1~mm thick and 35~mm effective diameter single-sided silicon detector (Micron MSD035) was placed around 10~cm downstream of the target to measure forward scattered protons and $\alpha$ particles. A sketch of the experimental setup is shown in the upper panel of Fig.~~\ref{fig:fig_1}. Particle identification was performed using the Time-of-Flight (about 1~ns resolution) and energy correlation, as shown in the lower panel of Fig.~\ref{fig:fig_1}. Elastic scattered protons are located in the low energy region of the plot, below 1000~keV, corresponding to 350~keV in CM. $\alpha$ particles coming from the decay of this particular resonance into $^{7}$Li+$^{4}$He would have an energy of about 4000~keV, quite far away from the region of interest. It is also worth pointing out that reactions on carbon atoms in the target that could produce $\alpha$ particles are highly suppressed at these bombarding energies due to the penetrability~\cite{Haider_1986}. The silicon detector was calibrated using a proton beam of different energies (down to 250~keV in the laboratory frame) and alpha particles from a $^{228}$Th source. The reaction energy was corrected by the energy loss of the particles in the target. The detector resolution of about 10$\pm$1~keV (FWHM) in the Center of Mass (CoM) was deduced taking into account the intrinsic resolution, straggling effects, and target thickness inhomogeneity. \\

\begin{figure}
\begin{center}
\includegraphics[width=\columnwidth, keepaspectratio]{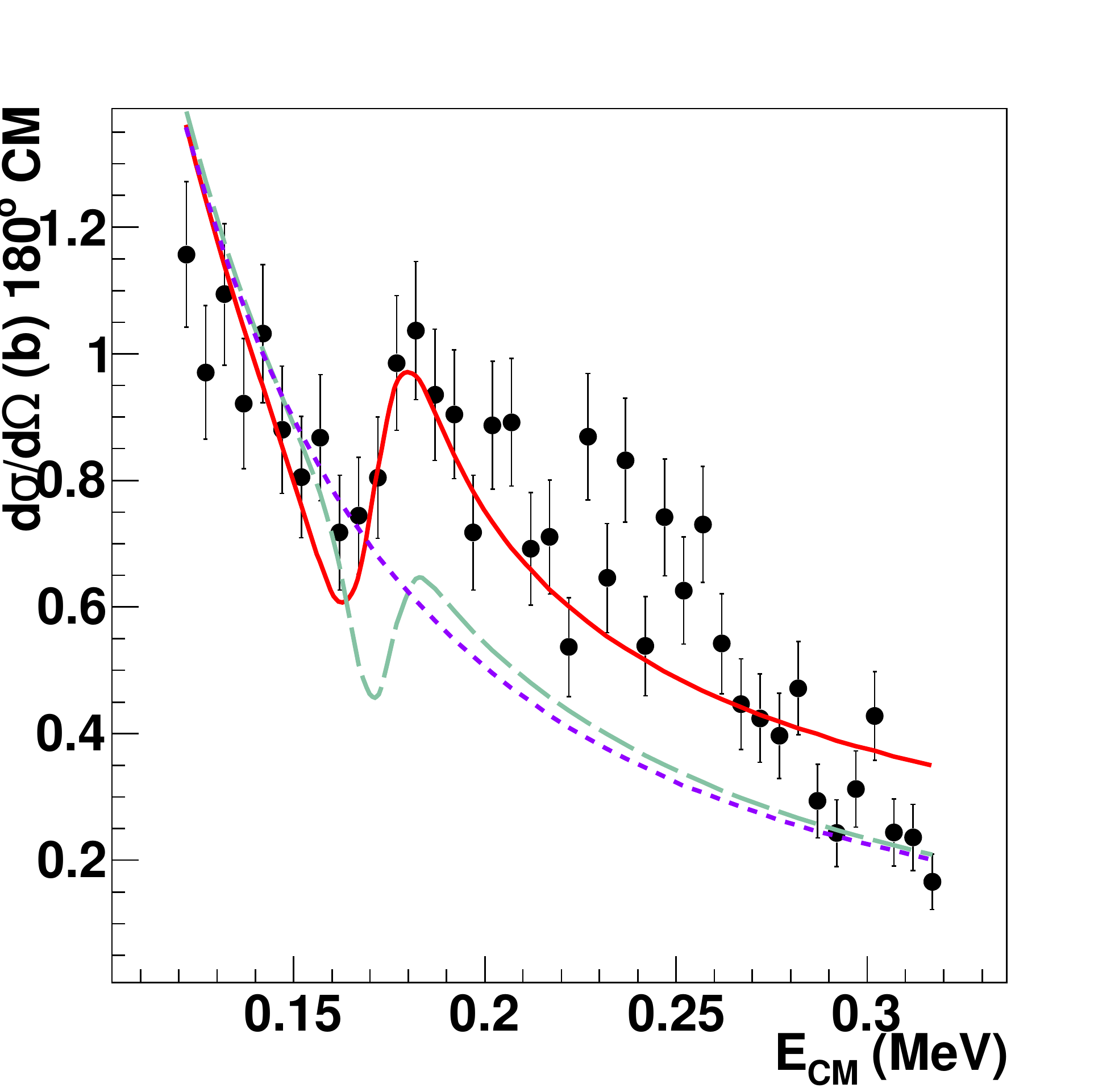}
\caption{\label{fig:fig_2} Excitation function (CoM) for the $^{10}$Be(p,p) reaction (solid dots) and best $R$-matrix fits performed  for 1/2$^{+}$ (solid line) and for 1/2$^{-}$  (dashed line).  The dotted line refers to the Coulomb scattering cross section.}
\end{center}
\end{figure}

Figure~\ref{fig:fig_2} shows the excitation function compared to an $R$-matrix calculation performed with the AZURE2 code~\cite{PhysRevC.81.045805}. These calculations were also compared to the ones yielded by the DSIGMAIV code~\cite{wang2006study}, finding an excellent agreement between the two codes. A resonance effect interfering with Coulomb scattering can be clearly seen below 200~keV. The resonance width ($16\pm3$~keV) and energy $E_R=171\pm20$~keV ($11.40\pm0.02$~MeV excitation energy in $^{11}$B) inferred from the fit are in good agreement with the values reported in Ref.~\cite{Ayyad2019} from the $^{11}$Be $\beta$-decay ($12\pm5$~keV). Moreover, the  best fit (solid line), with $\chi^{2}$=2.7, confirms the $J^{\pi}=1/2^{+}$ assignment. The proton partial width amounts to only $4.5\pm1.1$~keV. In order to obtain a resonance effect, compatible with the experimental resolution, another decay branch has been assumed and attributed to the $^{7}$Li+$\alpha$ decay channel ($11\pm3$~keV). The sharp energy cut at 350~keV in the CoM is due to the maximum beam energy. As it is evident from the figure, where the Coulomb scattering is also presented (dotted line), the cross section does not converge to Rutherford after the resonance. With such a narrow width, it would be expected that the cross section converges to pure Rutherford scattering if the resonance has a simple Breit-Wigner form. The excitation function exhibits a clear departure from the Breit-Wigner shape. Such deviations from the Breit-Wigner shape are well known~\cite{Cauldrons1988,Oliveira2018,PhysRevC.67.014308} and are mostly due to the energy dependence of the partial widths. Here the effect is enhanced by the interference effects with the (mostly Coulomb) background. An $R$-matrix fit for a resonance with $J^{\pi}=1/2^{-}$ ($\chi^{2}$=5.4) is also presented in Fig.~\ref{fig:fig_2} (dashed line). In this case, for $l=1$, the excitation function converges back to Coulomb scattering rapidly, in contrast to our data.\\

In order to corroborate that such a behavior is only due to specific properties inherent to the $R$-matrix formalism, such as the penetration factors, energy dependencies, resonance energy shifts and phase shifts~\cite{RevModPhys.30.257}, we performed a search of a potential resonance. For this, we employed the optical model code SPOMC~\cite{PhysRevC.66.014610}. The potential has been inferred from the optical model parameterization  of the code and renormalized to produce an $l=0, 2s$  resonance at the experimental energy. The imaginary potential was set to zero. The behavior obtained was very similar to the one with the $R$-matrix formalism, shown in Fig.~\ref{fig:fig_2}, with the cross-section remaining a about a factor of 2 higher than the potential scattering, even far from the resonance. Another excellent example of such a threshold change of the cross section due to the combined effect of interference and of the fast change of penetrability, similar to the one observed here, is a 1/2$^{+}$ low-lying resonance in $^{15}$F~\cite{DEGRANCEY2016}.\\

 This peculiar behaviour of the resonance effect can be described for a given partial wave by an interference between a potential scattering and the resonance term using the collision matrix for elastic scattering within the one resonance level  approximation~\cite{Descouvemont_2010}:

\begin{equation}
    U^{BW}_{cc'} = \exp{i(\phi_c+\phi_c') \bigg[ \delta_{cc'}} + \frac{i\sqrt{\Gamma_c(E)\Gamma_c'(E)} }{E_R-E-i\Gamma(E)/2} \bigg]
\end{equation}

 \noindent where the partial width is $\Gamma_{c} = 2\gamma^{2}_{obs,c}P_{c}(E)$, with $\gamma^{2}_{obs,c}$ and $P_{c}(E)$ being the observed reduced width and the penetrability, respectively. $\phi_{c}$ is the hard-sphere shift and $\delta_{cc'}$ is the Kronecker delta. In the case of a single channel, $\Gamma(E)$ corresponds to the width of the elastic channel. Below the Coulomb barrier, the penetrability may vary much faster than the term $(E_{R}-E)$. Then, the term $(E_{R}-E)$ can be neglected, and the expression will converge to $\delta_{cc'}-2$. This very specific behavior of  a near-threshold resonance has no longer a Breit-Wigner form, but resembles more a threshold behavior with a strong contribution to the scattering probability that remains almost constant after the resonance even at distances large as compared to the resonance width.\\

A measure of the single-proton content of the resonance is provided by the proton partial width, which can be extracted from the two-channel $R$-matrix fit of the excitation function, and it is proportional to the spectroscopic factor~\cite{RevModPhys.30.257,Descouvemont_2010}. A Wigner limit of the single particle width \cite{RevModPhys.30.257,Descouvemont_2010} for a channel radius of $2.7$~fm of 18~keV was obtained, yielding a spectroscopic factor of 0.25. This is in agreement with the conclusion of the previous work~\cite{Ayyad2019} suggesting that the resonance state contains a significant single-particle strength. Since the resonance is well below the Coulomb barrier, the experiment probes the asymptotic part of the corresponding state, and the elastic cross section is largely insensitive to the internal details of the wave function. The extracted spectroscopic factor is thus essentially proportional to the asymptotic normalization constant, which, as opposed to the spectroscopic factor, is an on-shell, well defined,  observable quantity. It is worth noting that in Ref.~\cite{Ayyad2019} the $\alpha$ decay could not be observed, due to the very strong branching to other channels decaying by $\alpha$ emission. Hence, the total width, as observed, contained the eventual contribution of this channel. In the present experiment, it was not possible to confirm directly the $\alpha$ decay, predicted to have around five times lower cross section, due to limited statistics. A direct measurement of the $^{10}$Be(p,$\alpha$), with a complete determination of the branching to different excited states in $^{7}$Li, is required to clarify the situation.\\

In conclusion, we have observed a near-threshold proton-emitting resonance in $^{11}$B via the $^{10}$Be(p,p) reaction at 350$A$~keV. An $R$-matrix calculation was used to deduce the energy, spin-parity, and resonance width, in good agreement with the values inferred in $\beta$-delayed proton emission of $^{11}$Be. This is a strong indication that the exotic $\beta p$ decay indeed proceeds via this intermediary state, explaining the relatively large branching ratio observed. The results also suggest that the resonance has a sizable decay width to the $\alpha$+$^{7}$Li channel. The characteristics of the resonance, a consequence of the interplay between the reaction mechanism and structure, reveals the open quantum system nature of such narrow resonances.\\

\begin{acknowledgments}

This work is based on the research supported in part  by the Spanish Ministerio de Ciencia, Innovación y Universidades. This work has received financial support from Xunta de Galicia (Centro singular de investigación de Galicia accreditation 2019-2022), by European Union ERDF, and by the “María de Maeztu” Units of Excellence program MDM-2016-0692 and the Spanish Research State Agency. Y.A. acknowledges the support by the Spanish Ministerio de Economía y Competitividad through the Programmes “Ramón y Cajal” with the grant number RYC2019-028438-I. This material is based upon work supported by the Department of Energy National Nuclear Security Administration through the Nuclear Science and Security Consortium under Award Number(s) DE-NA0003180 and/or DE-NA0000979, by the National Science Foundation, USA under Grant No. PHY-2012040, PHY-1565546, and PHY-1913554, and by  The U.S. Department of Energy, Office of Science, Office of Nuclear Physics under Grant DE-SC0020451 and contract DE-AC02-06CH11357. This work was performed under the auspices of the U.S. Department of Energy by Lawrence Livermore National Laboratory under Contract No. DE-AC52-07NA27344

\end{acknowledgments}

\bibliography{bibliography}

\end{document}